\documentclass[usenatbib]{mn2e}
\usepackage{epsfig}
\usepackage{rotating}
\usepackage{multirow}

\bibpunct[, ]{(}{)}{;}{a}{}{,}

\newcommand{\aj}{Astron. Journ.}

\newcommand{\pasp}{Publications of the Astronomical Society of the Pacific}

\title[Head-Tail Galaxies: Beacons of High-Density Regions in Clusters]{Head-Tail Galaxies: Beacons of High-Density Regions in Clusters}
\author[Mao et al.]{Minnie ~Y. Mao$^{1}$\thanks{E-mail: mymao@utas.edu.au}, Melanie Johnston-Hollitt$^{1,2}$, Jamie ~B. Stevens$^{1}$, 
\newauthor Simon ~J. Wotherspoon$^{1}$\\
$^{1}$School of Mathematics and Physics, University of Tasmania, Private Bag 37, Hobart, 7001, Australia\\
$^{2}$Excellence Cluster Universe, Technische Universit\"{a}t M\"{u}nchen,
Boltzmannstr. 2, D-85748, Garching, Germany}
\begin{document}

\date{2008}

\pagerange{\pageref{firstpage}--\pageref{lastpage}} \pubyear{2008}

\maketitle

\label{firstpage}

\begin{abstract}
Using radio data at 1.4 GHz from the ATCA we identify five head-tail (HT) galaxies in the central 
region of the Horologium-Reticulum Supercluster (HRS). Physical parameters of the HT galaxies were 
determined along with substructure in the HRS to probe the relationship between environment and 
radio properties. Using a density enhancement technique applied to 582 spectroscopic
measurements in the 2$^{\circ}$ $\times$ 2$^{\circ}$ region about A3125/A3128, 
we find all five HT galaxies reside in 
regions of extremely high density ($>$100 galaxies/Mpc$^3$). In fact, the environments surrounding HT galaxies are statistically denser than those environments surrounding non-HT galaxies and among the densest environments in a cluster. Additionally, the HT galaxies are found in regions of enhanced X-ray emission and we show that the enhanced density continues out to substructure groups of 10 members. We propose that it is the high densities that 
allow ram pressure to bend the HT galaxies as opposed to previously proposed mechanisms 
relying on exceptionally high peculiar velocities. 
\end{abstract}


\section{Introduction}
\label{htgalaxies}
The term ``radio galaxy'' is usually used to describe a class of objects which have characteristic 
jets of highly relativistic particles blasted out from the host galaxy. Although most if not all galaxies have some radio emission, the term ``radio galaxy'' is usually reserved for those galaxies that have a high radio-to-optical luminosity ratio \citep{Kellermann89}.

Head-tail (HT) galaxies are a sub-class of radio galaxy whose radio jets are distorted such that they appear to bend 
in a common direction. This gives rise to the so-called ``head-tail'' morphology where the radio jets 
are bent back to resemble a tail with the luminous galaxy as the ``head''. (The term ``head-tail'' was coined 
by \citet{Miley72} but the first HT galaxy was discovered by \citet{Ryle68}.) Although considered 
somewhat exotic in their early years, HT galaxies are now considered quite common. 

HT galaxies are generally detected in dynamical, non-relaxed clusters and are preferentially
found towards the centre of the cluster potential well \citep{Klamer04} or in regions of enhanced X-ray
emission indicative of a deep potential well \citep{Jones84}. Although it is common to find HT galaxies
in clusters, the number of HT galaxies per cluster is low with most unrelaxed clusters that 
have been probed yielding only one or two HT galaxies per cluster. 

HT galaxies are generally FRI in morphology \citep{Venturi98} which, assuming that the FRI/FRII break is caused by environment, is 
consistent with HT galaxies being found in clusters as their denser environments effectively restrict the outflow of the jets \citep{Owen94}. 

The bent morphology of jets in HT galaxies has always been attributed to one of two possible mechanisms
relating the relative velocity of the host galaxy to the intra-cluster medium (ICM). In the first 
mechanism the host galaxy has a larger than expected peculiar velocity, leading to ram pressure 
on the jets \citep{Miley72, Rudnick76}, where ram pressure, $P_{ram}$, is defined as:

\begin{equation}
P_{ram} \propto \rho v^2,
\end{equation}
where
$\rho$ is the density of the ICM, and 
$v$ is the relative velocity between the galaxy and the ICM \citep{Gunn72}. 

In the second mechanism, strong winds in the ICM, believed to be caused by dynamical interactions
such as cluster-cluster mergers or accretion of groups on to clusters, bend the jets. The phenomenon
is often referred to as ``cluster weather'' \citep{Burns98}. These mechanisms have been
invoked to explain the difference in the degree to which the jets are bent, with Narrow-Angle Tailed galaxies
(NATs) believed to be formed via ram pressure in cases where the host galaxy has a peculiar
velocity of the order of 600 km s$^{-1}$ and is embedded in a
hot, dense plasma \citep{Venkatesan94}, and Wide-Angle Tailed galaxies (WATs) believed to be the result of
``cluster weather'' \citep{Klamer04}. NATs are most likely to be caused by the passage of the galaxy through the cluster \citep{Odea85} while WATs are more likely to be associated with dominant cluster galaxies \citep{Owen76}.

Although the distinction between ram pressure and ICM effects, 
such as ``cluster weather'' \citep{Burns98}, 
is often drawn in literature, we note that cluster weather still implies ram pressure as the cause of the 
jet bending. The true distinction between the ram pressure theory and the cluster weather theory lies in 
the velocities of the HT galaxies relative to the kinematic centre. Henceforth, when we refer to the ram pressure 
theory we are indicating that the galaxy has a high peculiar velocity and is in fact travelling at a 
velocity which differs significantly from the statistical distribution of velocities for other galaxies 
in the cluster. Cluster weather on the other hand will mean that the galaxy does not have a 
significantly different velocity, but rather it is the motion of the ICM that is causing the 
bent morphology. 

In both cases it is the relative motion of the host galaxy with respect to the ICM that causes the 
bending. Consequently when only viewing the radio morphology it is not possible to distinguish 
between a galaxy with high peculiar velocity 
moving through an undynamical, stationary ICM, and a galaxy with the ICM rushing past. However, 
spectroscopic redshifts can be used to determine the distribution of velocities for cluster galaxies 
and, via a statistical analysis, the relative importance of the host galaxy's motion with respect to 
its neighbours can be established. 
 
This paper explores the environment of a sample of HT galaxies and examines their usefulness as a 
tool to investigate dynamical conditions in the host clusters. We examine the 
well-studied major merger A3125/A3128 which resides in the central region of the Horologium-Reticulum 
Supercluster (HRS). Using a large spectroscopic dataset in conjunction with multi-frequency radio imaging
we present a quantitative analysis of the environment of HT galaxies as compared to samples
of both other cluster galaxies and other cluster radio galaxies. The paper is laid out as follows:
Section 2 discusses the spectroscopic and radio data in the region surrounding A3125/A3128 and the parameters
associated with the detected HT galaxies; Section 3 presents the substructure detection; Section 4
gives the results of our statistical analysis of the environments surrounding HT galaxies and Sections 5 and 6 present the discussion and conclusions.

This paper uses H$_0$ = 70 km s$^{-1}$ Mpc$^{-1}$ and $\Omega$$_M$ = 0.27. This gives a 
scale of 1.191 kpc/arcsecond at 18500 km s$^{-1}$, the mean redshift of the HRS. 

\section{Head-Tail Galaxies: Detection and Morphology}
To investigate the relationship between HT galaxies and the dynamics of the cluster, we study HT galaxies in A3125/A3128, a major merger in the central region of the Horologium-Reticulum supercluster (HRS). The HRS is the second largest supercluster in the local universe (out to 300 Mpc), preceded only by the Shapley supercluster. The cores of rich superclusters are ideal environments to study cluster mergers because the high peculiar velocities induced by the enhanced local density favours cluster-cluster and cluster-group collisions \citep{Venturi00}.

\subsection{Spectroscopic Data}
We compiled a catalogue of 2181 spectroscopic redshifts in the central region of the HRS. Data were obtained from all previously published surveys \citep{Rose02, Fleenor06} in addition to an unpublished catalogue of spectroscopic data obtained at ANU by T. Mathams in 1989 and with data from the NASA Extragalactic Database (NED). A spatially and velocity limited subset of the 2181 galaxies covering 2$^\circ$ $\times$  2$^\circ$ in area and 15,000 km s $^{-1}$ $\leq$ cz $\leq$ 22,000 km s$^{-1}$ in velocity centred on the major merger A3125/A3128 yields 582 redshifts.

When compiling a catalogue from several different sources which have overlapping observations of the same galaxies, care must be taken to ensure that individual objects do not appear more than once in the final catalogue or that close, but independent objects are not incorrectly conflated. To avoid these problems we employ a visual inspection method for all cases where objects from different sources appear within the spatial and recessional uncertainties of the catalogues 18$''$ and 400 km s$^{-1}$). In such cases postage stamps from the Digitized Sky Survey (DSS) are inspected. This is an important step as a completely automated compilation method might inadvertantly remove true close-proximity galaxies.

\begin{figure*}
\begin{center}
\begin{tabular}{cc}
\includegraphics[scale=0.5]{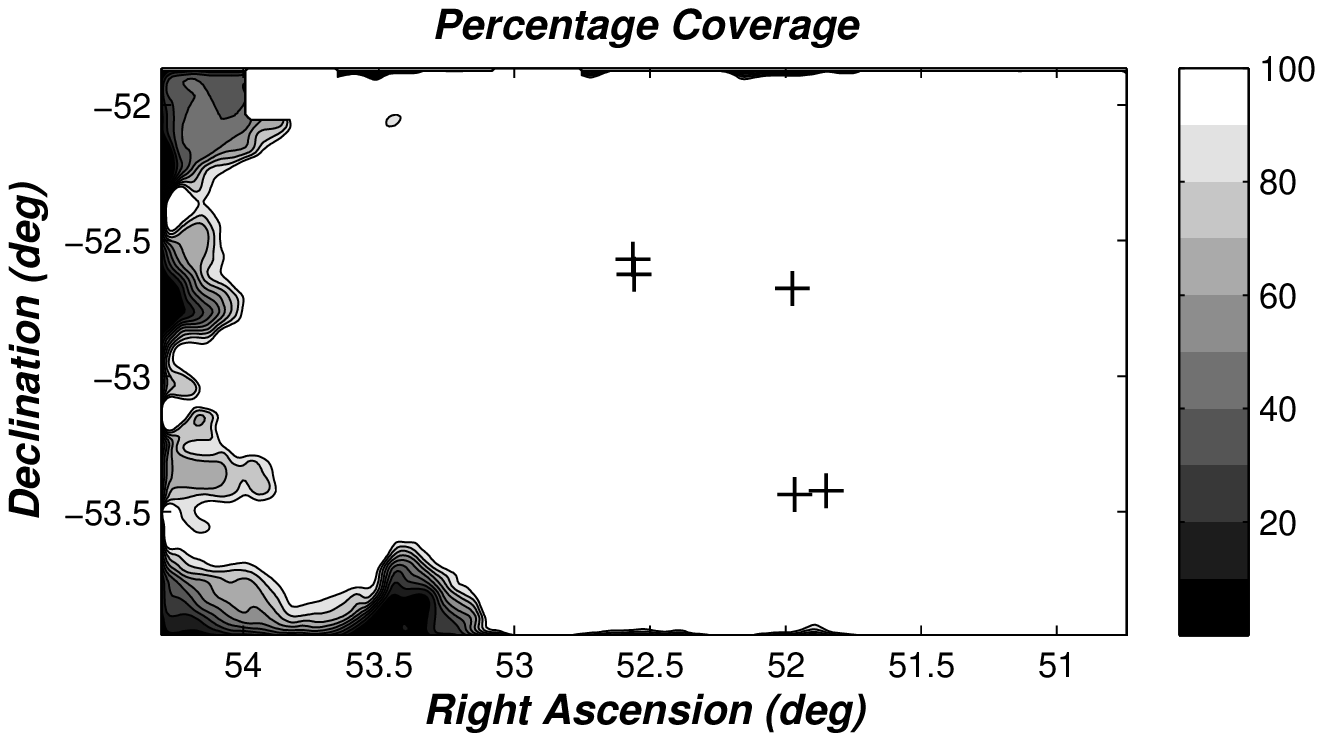} & \includegraphics[scale=0.5]{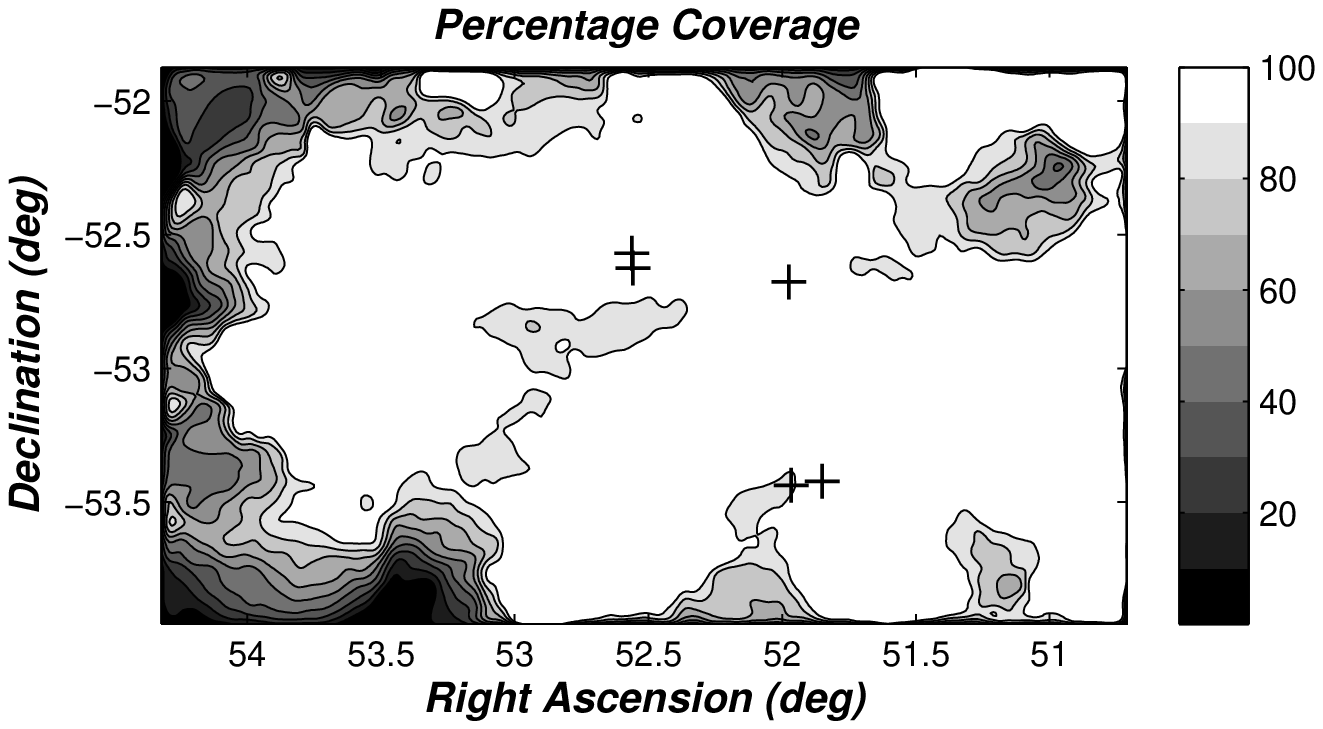}\\
\includegraphics[scale=0.5]{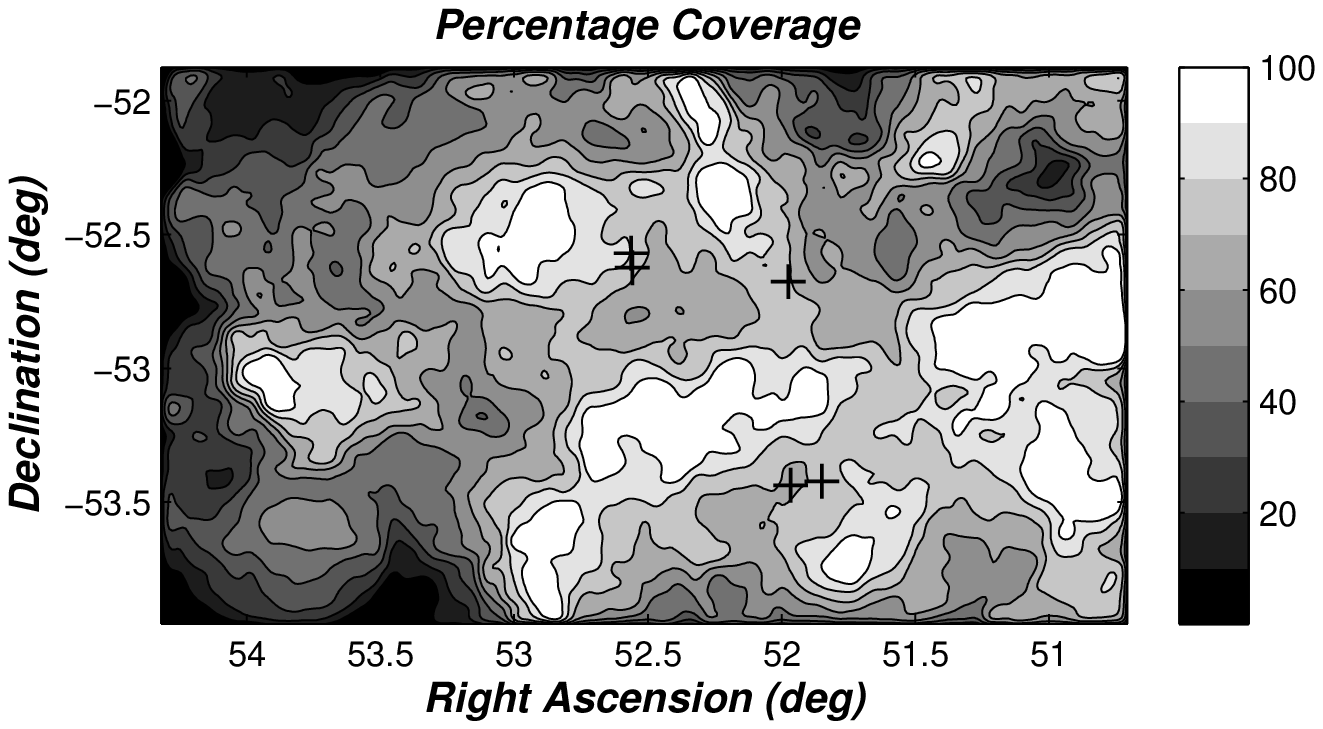} & \includegraphics[scale=0.5]{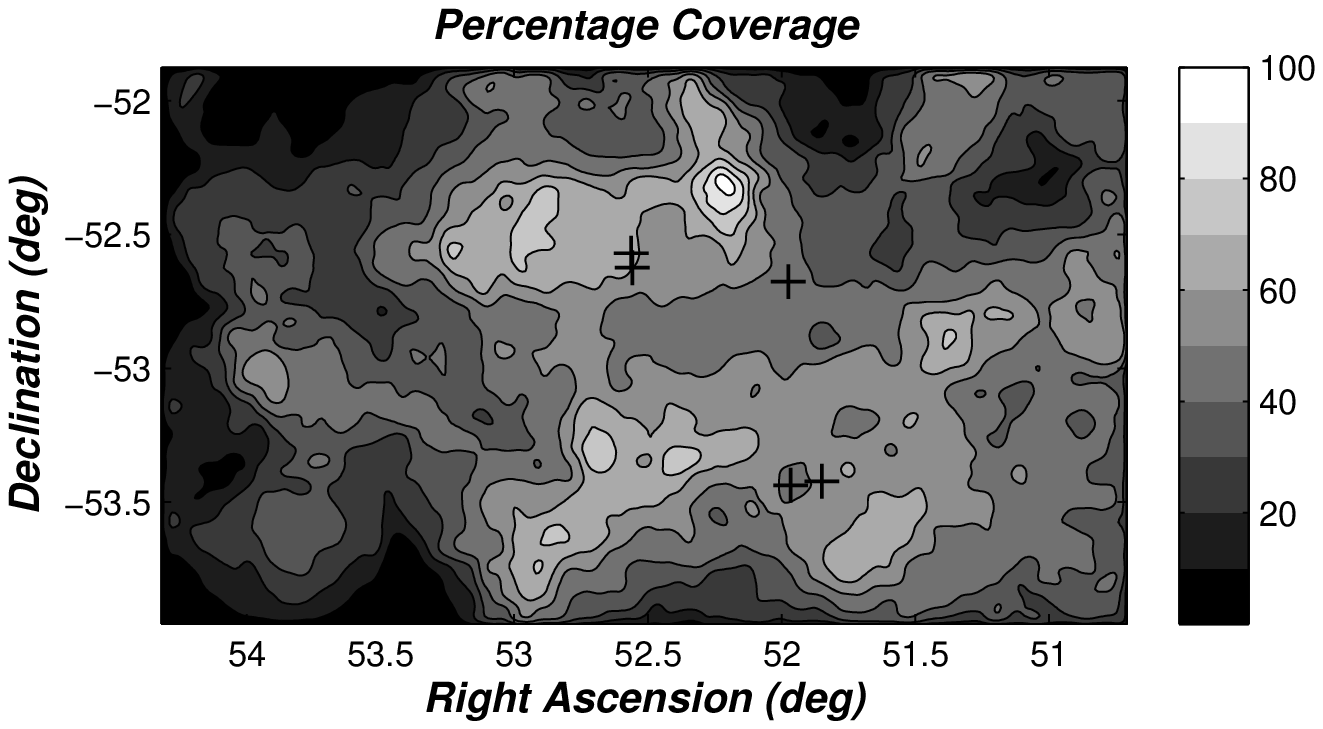}\\
\end{tabular}
\end{center}
\caption {Top left: Skyplot of the HRS region with m$_b$ cutoff = 18. Top right: Skyplot of the HRS region with m$_b$ cutoff = 18.5. Bottom left: Skyplot of the HRS region with m$_b$ cutoff = 19. Bottom right: Skyplot of the HRS region with m$_b$ cutoff = 19.5. Contours are set at 10\% intervals. The crosses mark the positions of the HT galaxies.}\label{skyplot}
\end{figure*}

The final catalogue of 582 galaxies has a magnitude limit of m$_b$=23 and is $\sim$90\% complete to m$_b$=18 and $\sim$60\% complete to m$_b$=19.5. The optical coverage of the HRS region appears uniform (see Figure \ref{skyplot}).

\begin{figure*}
\begin{center}
\includegraphics[angle=-90, scale=0.5]{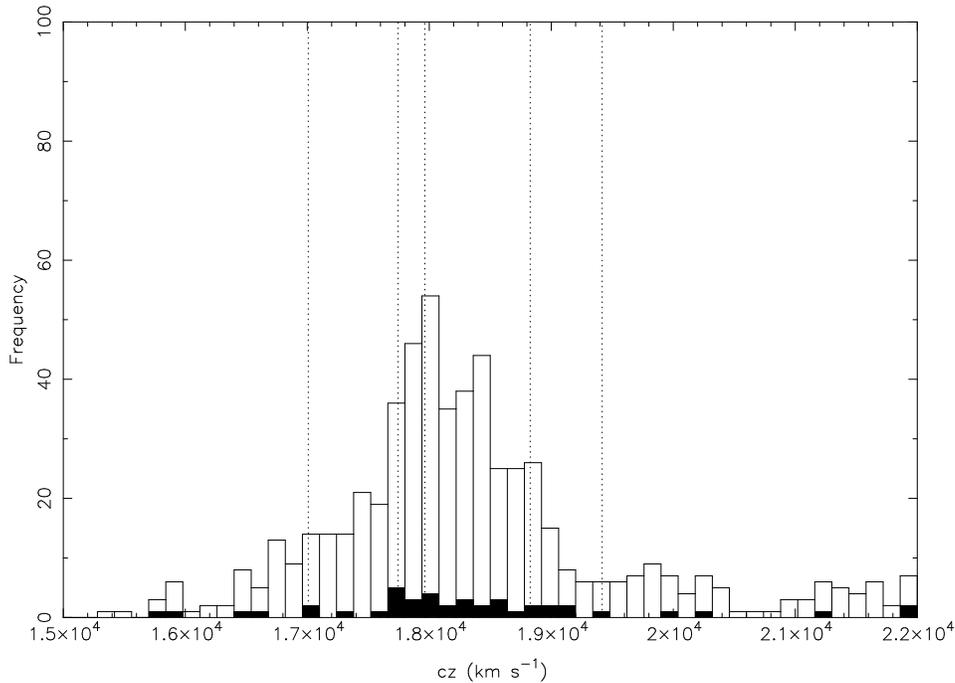}
\end{center}
\caption{Histogram of galaxy velocities from our position and velocity limited catalogue (of 582 galaxies).The shaded histogram plots the 46 radio galaxies. The velocities of our HT galaxies are shown with dotted lines. It can be seen that the HT galaxies are clearly in the main component of A3125/A3128, as are the radio galaxies.}\label{cat_vel}
\end{figure*}

The resultant spectroscopic catalogue, when combined with the available radio images, provides an ideal dataset with which to investigate the environment of HT galaxies. Figure \ref{cat_vel} shows the distribution of velocities of our catalogue.

\subsection{Radio Observations}

\begin{figure*}
\vspace{-2cm}
\begin{center}
\includegraphics[width=15cm, bb=0 0 600 600]{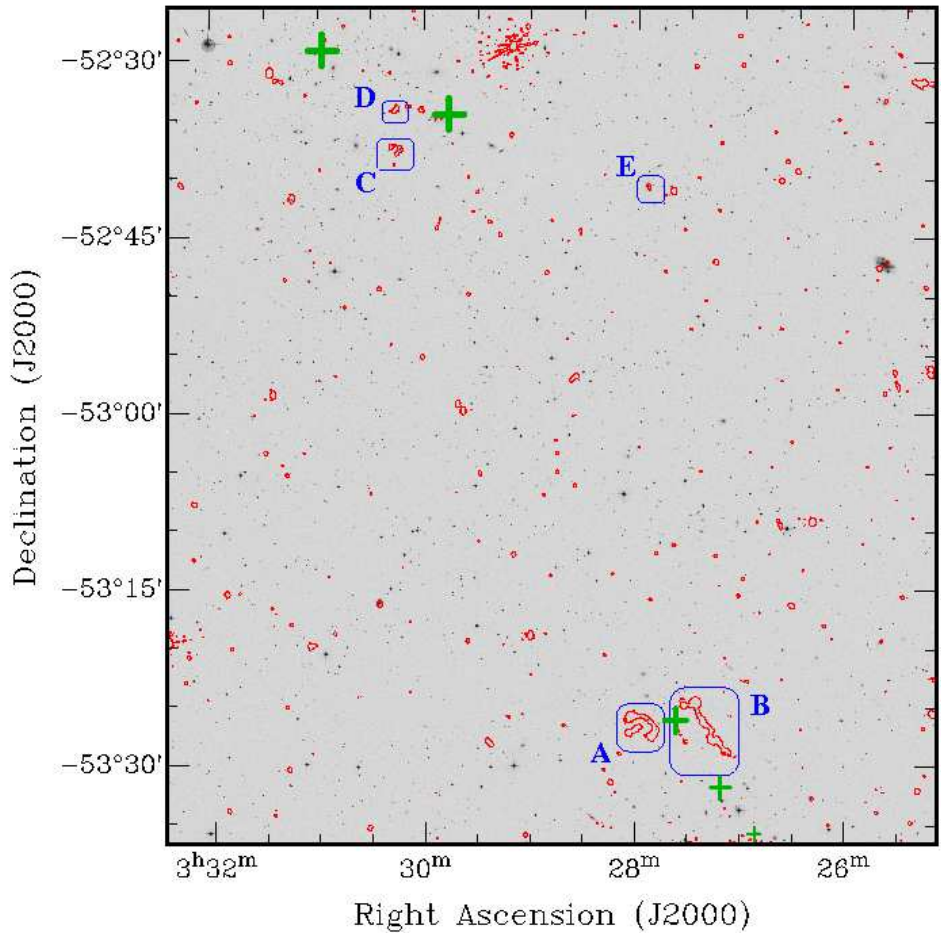}
\end{center}
\caption{Central region of the 3.28 square degrees 1.4 GHz image of A3125/A3128 overlaid on optical data from the DSS. The RMS of the image is 0.1 mJy beam$^{-1}$ and the contour is at three times the RMS noise. The five HT galaxies identified in the image are denoted with boxes. We note that Galaxy A appears to have a companion galaxy nestled between its lobes. This galaxy is in fact in the background and hence not considered in this study. The X-ray peaks of A3125 (lower right with three peaks) and A3128 (upper left with two peaks) are marked with crosses - the size indicates the strength of the X-ray emission.}\label{radiowhole}
\end{figure*}

Radio images at 1.4 GHz of the central 3.28 square degrees surrounding the A3125/A3128 system 
(centred at 03:29:45, -53:00:00) were obtained with the Australia Telescope Compact Array (ATCA). 
The observation was a 40-pointing, linear mosaic consisted of 2 $\times$ 128 MHz observations centred 
at 1344 and 1432 MHz. The data were obtained in three different array configurations over 39 hours 
in 2001 (see Table \ref{config}). The primary calibrator was PKS B1934-638 and the phase calibrator 
was PKS B0302-623. The synthesized beam size is 9.90 by 7.87 arcseconds with a position angle of 31.9$^\circ$. 
The mean RMS noise for the image (Figure~\ref{radiowhole}) is 0.1 mJy beam$^{-1}$ and lowest 
contours are set at 0.3 mJy beam$^{-1}$, three times the RMS noise. Over 500 radio galaxies are 
detected in the 1.4 GHz data \citep{mjh_rlf08} of which 46 are identified with optical galaxies 
having redshifts found in our spectroscopic catalogue that placed them in A3125/A3128.

\begin{table*}
\begin{center}
\caption{Details of the 1.4 GHz ATCA observations.  Columns 1 and 2 give the telescope configuration and observing date respectively. Column 3 is the integration time in minutes, column 4 is bandwidth in MHz, column 5 gives
the central frequencies for each IF and column 6 is the uv-range covered
in kilowavelengths. 
Observations were carried out by R.~W. Hunstead,
M. Johnston-Hollitt, J.~A. Rose, W. Christiansen, \& M. Fleenor. Full details of the
observations and reduction are available in \citep{mjh_rlf08}.}
\begin{tabular}{llllll}
\\
\hline
\hline
Configuration & Date & Integration & Bandwidth & Frequencies & UV-range\\
&&(min)& (MHz)& (MHz) & (k$\lambda$)\\
\hline
6A & 8th \& 10th December, 2001 & 880 & 2 $\times$ 128 & 1344 \& 1432 & 0.50 - 20.72 \\
6D & 27th November, 2001 & 840 & 2 $\times$ 128 & 1344 \& 1432 & 1.57 - 27.72\\
1.5D & 19th \& 20th November, 2001 & 640& 2 $\times$ 128 & 1344 \& 1432 & 0.36 - 27.43\\
\hline
\hline
\label{config}
\end{tabular}
\end{center}
\end{table*}

Low frequency data for the region were also available from the Sydney University Molonglo Sky Survey (SUMSS) \citep{Mauch03}, 
which used the Molonglo Observatory Synthesis Telescope (MOST) to assemble a comprehensive picture 
of the Southern radio sky ($\delta \leq -30^\circ$) at 843 MHz. Data obtained with the MOST have a 3 MHz bandwidth and the
resultant images from SUMSS have a resolution of the order of 1$'$.

\subsection{Identifying HT Galaxies}

Five HT galaxies are identified in the 1.4 GHz data, including two that were previously 
detected \citep{mjh04}. Typically only one or two HT galaxies tend to be found in non-relaxed 
clusters in the current literature. However, the available radio surveys covering clusters tend either to be quite shallow, or have low sensitivity to extended emission (e.g. the NVSS \citep{Condon98}) and hence only 
the brighter extended sources are detected. In a comparably shallow survey to those commonly used we would
 only be able to see the two HT galaxies previously studied by \citet{mjh04} as they are the brightest 
HT galaxies of our sample. Interestingly, \citet{Miller05} surveyed 7 square degrees in the central 
region of the Shapley Supercluster to a similar depth to our data (80 $\mu$Jy beam$^{-1}$ at a resolution of 16$''$), and also found five HT galaxies. 

The optical and radio parameters of the HT galaxies are discussed below and Table \ref{opticalparameters} summarises the respective frequency data.

\begin{table*}
\begin{center}
\caption{Optical and radio parameters of the HT galaxies. Column 1 is the identifier of the HT galaxy while column 2 gives its NED ID. Columns 3 and 4 give the Right Ascension and Declination in J2000 coordinates while columns 5, 6, 7 and 8 give the redshift, luminosity distance \citep{Wright06}, apparent magnitude and absolute magnitude respectively. All magnitudes are b$_j$, except for galaxy E which is B$_T$. Columns 9 and 10 give the flux density and power of the galaxies at 1.4 GHz respectively and column 11 gives the Fanaroff-Riley morphology class.}
\tiny{
\begin{tabular}{llllllllcll}
\\
\hline
\hline
\multirow {2}{*}{ID} & \multirow {2}{*}{Optical Counterpart} & RA & Dec & cz & D$_{\rm lum}$ &\multirow {2}{*}{mag} &\multirow {2}{*}{Mag} & Flux$_{1.4}$ & Power$_{1.4}$ & \multirow {2}{*}{Type}\\
& & (J2000) & (J2000) & (km s$^{-1}$) & (Mpc) & & & (mJy) & (W/Hz) & \\
\hline
A & 2MASX J03275206-5326099 & 03 27 52 & -53 26 10 & 17745 & 265.0 & 16.20 & -20.92 & 245 $\pm$ 25 & 1.84 $\times$ 10$^{24}$ & FRI/II\\
B & 2MASX J03272476-5325179 & 03 27 25 & -53 25 18 & 18827 & 282.0 & 15.85 & -21.40 & 529 $\pm$ 53 & 4.46 $\times$ 10$^{24}$ & FRI/II\\
C & 2MASX J03301366-5237304 & 03 30 14 & -52 37 28 & 19417 & 291.2 & 16.24 & -21.08 & 35 $\pm$ 4 & 3.12 $\times$ 10$^{24}$ & FRI\\
D & 2MASX J03301522-5234124 & 03 30 15 & -52 34 10 & 17008 & 253.6 & 16.94 & -20.08 & 24 $\pm$ 2 & 1.62 $\times$ 10$^{24}$ & FRI\\
E & IC 1942 NED01 & 03 27 54 & -52 40 35 & 17964 & 268.5 & 17.27 & -19.87 & 11 $\pm$ 1 & 8.37 $\times$ 10$^{24}$ & FRI\\
\hline
\hline
\label{opticalparameters}
\end{tabular}
}
\end{center}
\end{table*}

\subsubsection{Optical Counterparts}
The positional data for the HT galaxies is taken as the coordinates of their optical counterparts; these values were obtained from NED. The HT galaxies all lie at comparable distances and have similar magnitudes. The velocities of the HT galaxies are marked by dotted lines in Figure \ref{cat_vel}.

\subsubsection{Flux Densities}
We used the Karma \citep{Gooch96} package Kvis to determine the flux densities of the HT galaxies at both available frequencies. The error in the flux density is 

\begin{equation}
\Delta S = \sqrt{(Sb)^2 + a^2},
\end{equation}
where $S$ is the flux density,
$b$ is the residual calibration error, and
$a$ is the RMS noise.

Because the flux density of our radio galaxies is so high relative to the RMS noise, the uncertainty in total flux density is dominated by the 10\% residual calibration error typical for the ATCA.

\subsubsection{Power and Fanaroff-Riley Classification}

Using the radio power formula from \citet{Hogg99} we calculate the radio powers for the five HT galaxies. Their powers range between $8.37 \times 10^{22}$ W Hz$^{-1}$ and $4.46 \times 10^{24}$ W Hz$^{-1}$. We find all of our HT galaxies to be either FRI or at the transition between FRI and FRII, which is consistent given the absolute magnitudes of these galaxies \citep{Owen00}. This is in agreement with previous literature \citep[eg.][]{Venturi98}.

\section{Substructure Detection}
\label{substructure}
 Substructure, which may be defined as the presence of two or more clumps of matter within a cluster, is a clear sign of incomplete relaxation and is hence indicative of the dynamical nature of a cluster of galaxies \citep{Pinkney96}. 

Substructure can be detected in two-dimensions via analysis of X-ray surface brightness data, but to investigate the full three-dimensional nature of a cluster a large spectroscopic survey is required.
 
By identifying substructure in the cluster we can investigate the relationship between the location of the HT galaxies and the underlying dynamical structure. 

\citet{Caldwell97} showed that the A3125/A3128 system is in a post-merger state and that A3125, the smaller of the two clusters, is more dispersed due to its recent passage through A3128. 

Substructure in the central region of the HRS has been studied previously by \citet{Rose02}. Using their spectroscopic data for 532 galaxies in this region, \citet{Rose02} identified four groups on the high velocity side of A3125/A3128. These groups were selected on the basis of their clumping in either RA-cz space or Dec-cz space. We note that in this context ``groups'' refer to substructure groups and not galaxy groups. 

Furthermore, \citet{Rose02} noted the presence of a void 4000 km s$^{-1}$ on either side of the clusters. They interpret this as a consequence of the large gravitational potential of the cluster and hence the voids are indicative of galaxies being absorbed from the peripheries into the clusters. 

The substructure detection method used by \citet{Rose02} is limited in that groups can only be oriented along the RA, Dec or cz axes. This means any ``infalling groups'' that are falling ``diagonally into the cluster'', for example, will not be as easily detectable.

We believe that rather than identifying substructure in the A3125/A3128 region, \citet{Rose02} 
have merely identified parts of the cluster. For this reason we have chosen not to repeat the 
method used by \citet{Rose02} when detecting substructure. Instead we use density enhancements in our catalogue to 
determine the location of substructure in this region. Using a physical parameter such as density
enhancement provides a more accurate analysis of the substructure within A3128/A3125.

\subsection{3D Density Enhancements}
To identify the location of substructure in A3125/A3128 we introduce our own method of identifying substructure using density enhancements.

We searched for density enhancements in our catalogue of galaxies using the HIPASS group-finder, a group
finding program written by \citet{Stevens05}\footnote{http://www-ra.phys.utas.edu.au/~jstevens/research.html}. This group finder needs only sky
position and velocity to operate, and finds groups using the hierarchical algorithm \citep{Tully87}.

Specifically, the group-finder uses the grouping equations of \citet{Gourgoulhon92} (hereafter GCF92) to determine the
likely three-dimensional separation of two galaxies using their on-sky positions and recessional
velocities. The GCF92 equations compute separation in three regimes:
\begin{enumerate}
 \item {When the velocities of the two galaxies differ by less than some user-specified parameter
$V_{l}$, the velocity difference is considered to be entirely peculiar, and thus the separation is
calculated using only plane-of-sky information.}
\item {When the velocities of the two galaxies differ by more than some user-specified parameter $V_{R}$,
then the galaxies are considered to be ungroupable; ie. they are separated by Hubble expansion and are
not likely to be bound to each other.}
\item {When the velocities of the two galaxies differ by an amount between these two extremes,
then the separation is calculated using a combination of plane-of-sky and velocity information.}
\end{enumerate}

At the start of the procedure, each galaxy is considered as an individual ``unit''. The group-finder
calculates the separation between every pair of units using the GCF92 equations, and from this the
radius of the groups (the average galaxy-galaxy separation of the group members) that would result from each pair. The galaxy number density for each potential
group is then calculated simply as the total number of galaxies in the two units divided by the volume. This
differs from the method of GCF92, who calculated mass density, but this is the only
significant difference.

The pair of units with the highest number density are then grouped into a single unit, which replaces
the two original units. The separations between each of the remaining pairs of units in the
catalogue are then recalculated and the process begins again. It should be noted that a unit with more
than one galaxy does not hide any information: when the radius of a group is calculated, the group-finder
always considers the individual galaxies. The group-finder stops when the number density of the densest
pair is below some user-specified stopping density. For each group that it identifies, the group-finder
calculates both the radius as calculated from the GCF92 equations and the projected radius,
using just the plane-of-sky information.

Because the group-finder uses number density to identify groups, it can be thought of as a type of
nearest-neighbour finder. Indeed, if $V_{l}$ is set to a sufficiently large value, and $V_{R}$ was set
to the same value, the group-finder would use only plane-of-sky information and would identify only
those groups which appear as dense regions on the sky. However setting $V_{l}$ lower will allow
the group-finder to identify structure along the line of sight.

A comparison could be made to the non-parametric algorithm DEDICA \citep{Pisani93,Pisani96}, which was designed
to identify clusters and determine whether they have substructure. However, the work presented in this paper
is less concerned with the identification of the clusters themselves, and more concerned with the environment
immediately surrounding each galaxy in the cluster. As the hierarchical algorithm gives information about
clustering strength at all scales, it is particularly suited for our purposes. 

\begin{figure}
\begin{center}
\includegraphics[angle=-90, scale=0.3]{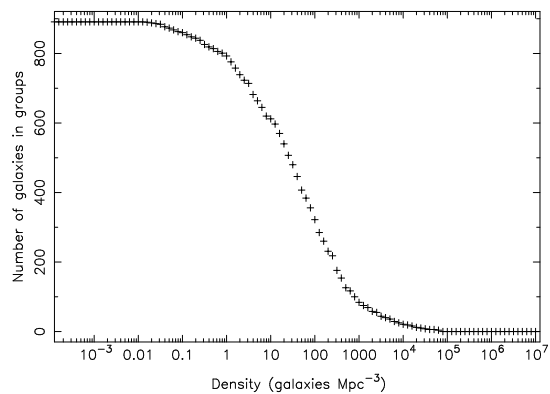}
\end{center}
\caption{Results of the HIPASS group finder at $V_l$ = 1000 km s$^{-1}$. The x-axis gives the density in galaxies/Mpc$^3$ on a logarithmic scale while the y-axis is the number of galaxies that can be found in groups.}\label{vl1000}
\end{figure}

\subsection{2D Density Enhancements}
To minimize the uncertainties that accumulate when using velocity as a pseudo-spatial parameter, we repeat the density enhancement study using projected radii instead of 3D density. That is, the substructure groups are still selected using exactly the same method as above, but we measure the projected radii of the groups to determine relatively whether HT galaxies reside in different environments to non-HT galaxies.

For each galaxy in the catalogue, as more galaxies are grouped with it, we can track how the projected radii increases as well, in effect showing us how density changes with increasing group membership.

\section{Results}
From Figure \ref{cat_vel} we can deduce that all the HT galaxies lie in the main peak of the velocity distribution. Thus it appears that the HT morphology is not a consequence of high peculiar velocity. 

Figure \ref{vl1000} shows the fraction of the spectroscopic catalogue assigned to a group as a function of stopping density. It is clear that  at very low densities ($\sim$10$^{-2}$ galaxies/Mpc$^3$), all the galaxies in the catalogue are placed in groups and  at very high densities ($\sim$10$^{5}$ galaxies/Mpc$^3$) almost no galaxies are in groups as expected.

The cut-off densities of the HT galaxies-- that is, the maximum density of the groups containing a HT galaxy -- show that HT galaxies reside in extremely dense regions. Table \ref{HTdensity} gives the cut-off densities for each of the HT galaxies. We note that due to incomplete optical coverage of our region not all redshifts have been determined hence the cut-off densities are lower limits. This is especially true for Galaxies A and C where there clearly exists close galaxies, but without reliable redshifts these galaxies are not incorporated in our cut-off density calculation.

\begin{table}
\begin{center}
\caption{Cut-off densities for HT galaxies.}\label{HTdensity}
\begin{tabular}{lll}
\\
\hline
\hline
& $V_l$ = 1000 km s$^{-1}$\\
HT galaxy & Cut-off density\\ & (galaxies/Mpc$^3$)\\
\hline
A & 645\\
B & 60680\\
C & 135\\
D & 20072\\
E & 64919\\
\hline
\hline
\end{tabular}
\end{center}
\end{table}

In fact, we find that the HT galaxies are in the densest regions of the cluster. We assume that the density of the environment has no bearing on whether a galaxy is radio-emitting (as per \citet{Venturi00}).

\subsection{Density Cut-Offs}
To ensure that the extremely dense environments surrounding our HT galaxies are statistically significant, we perform Monte Carlo shuffles on our density cut-offs.

Monte Carlo shuffles work by calculating the mean (or median) density cut-off of the five HT galaxies. We then choose five random galaxies and average their density cut-offs. This is repeated 100000 times and the fraction of shuffles that produces density cut-offs greater than the mean (or median) density cut-off of the HT galaxies show the statistical significance of the HT galaxies' density cut-offs.

If HT galaxies are truly in denser regions than non-HT galaxies we would expect that 
\begin{itemize}
\item{HT galaxies are found in statistically denser regions than non-HT galaxies with radio emission,}
\item{HT galaxies may or may not be found in statistically denser regions than non-HT galaxies without radio emission. This is because galaxies without radio emission can potentially be in denser regions than HT galaxies but as they have no radio morphology, a comparison is meaningless, and}
\item{radio galaxies are not found in statistically different regions to non-radio galaxies \citep{Venturi00}.}
\end{itemize}

Consequently, we perform Monte Carlo shuffles on three different sets of sample populations.
\begin{enumerate}
\item{HT v non-HT radio}
\item{HT v non-HT}
\item{radio v non-radio}
\end{enumerate}

The results are shown in Table \ref{pvalue}. A result is considered statistically significant when the Monte Carlo fraction is less than or equal to 0.01.

\begin{table}
\begin{center}
\caption{Fraction of Monte Carlo Shuffles that have cut-off densities greater than or equal to the mean (or median) HT cut-off density.}
\scriptsize{
\begin{tabular}{lll}
\\
\hline
\hline
Test & Mean & Median\\
\hline
HT v non-HT radio & $<$0.001 & 0.001\\
HT v non-HT & $<$0.001 & $<$0.001\\
non-radio v radio & 0.011 & 0.081\\
\hline
\hline
\label{pvalue}
\end{tabular}
}
\end{center}
\end{table}

The values generated for HT v non-HT radio and HT v non-HT are statistically significant for both average and median statistics which means that HT galaxies do reside in regions of enhanced density. Using the mean, it appears that there is a significant difference between radio and non-radio galaxies, but this significance disappears when using the median statistic. We attribute this to the very high densities of the HT galaxies, which could skew the radio mean, but would have little effect on the median.

\subsection{Radii}

To verify the results from the density parameter, and to explore the environment further out from each of the HT galaxies, we conduct an analysis using projected radius. For this, the groups are still identified using the HIPASS group finder in the usual way, however we now examine their projected radii as the number of galaxies in the group (group population) increases. 

Table~\ref{HTradii} gives the projected radii of the groups found around each HT galaxy, while Table~\ref{averadii} shows the average projected radii of the groups found around all the galaxies in the catalogue, and is broken up according to galaxy class (ie. HT, radio, all). The average group radii for HT galaxy groups is roughly half that of those around other classes of galaxies, for groups of less than 10 members. This suggests that HT galaxies are surrounded by closer companions than non-HT galaxies, and supports the result that HT galaxies reside in denser regions of the cluster.

\begin{table}
\begin{center}
\caption{Radii of the groups containing HT galaxies. Column 1 is the HT galaxy and the remaining columns give the projected radius of the group (in kpc) with the specified population.}\label{HTradii}
\tiny{
\begin{tabular}{crrrrrrrr}
\\
\hline
\hline
HT galaxy & \multicolumn{8}{c}{Group population}\\
 & 2 & 3 & 4 & 5 & 10 & 15 & 20 & 30\\
\hline
A & 71 & 126 & 174 & 223 & 402 & 540 & 848 & 1092 \\
B & 16 & 29 & 43 & 93 & 264 & 357 & 558 & 860\\
C & 119 & 139 & 158 & 177 & 273 & 360 & 474 & 1172\\
D & 23 & 53 & 105 & 134 & 392 & 591 & 721 & 1222\\
E & 15 & 81 & 108 & 141 & 296 & 517 & 811 & 1127\\
\hline
\hline
\end{tabular}
}
\end{center}
\end{table}

\begin{table}
\begin{center}
\caption{Average radii for all groups. Column 1 is the class of galaxy (HT: the five HT galaxies; radio: the 46 radio galaxies (including the five HT galaxies); all), and the remaining columns give the average radii of the groups (in kpc) with the specified population.}\label{averadii}
\tiny{
\begin{tabular}{crrrrrrrr}
\\
\hline
\hline
Class & \multicolumn{8}{c}{Group population}\\
 & 2 & 3 & 4 & 5 & 10 & 15 & 20 & 30\\
\hline
HT & 49 & 86 & 118 & 154 & 325 & 473 & 682 & 1095\\ 
Radio & 95 & 153 & 211 & 261 & 504 & 674 & 751 & 1055\\
All & 109 & 163 & 216 & 268 & 489 & 645 & 749 & 1020\\
\hline
\hline
\end{tabular}
}
\end{center}
\end{table}

We perform Monte Carlo shuffles on our data to determine whether the smaller radii associated with HT galaxy groups are significant. We calculate the average radius of the HT galaxy groups as a function of group population, and then choose five random galaxies and calculate their average group radii in the same way. This is repeated 100000 times, and the fraction of shuffles that produce an average radius smaller than the average HT galaxy group radius for the same group population is shown in Table \ref{monteradii}.

\begin{table}
\begin{center}
\caption{The fraction of Monte Carlo Shuffles that produced five groups with an average radius less than the average radius of the HT galaxy groups, when all groups are considered and when only groups around radio galaxies are considered.}
\begin{tabular}{crr}
\\
\hline
\hline\
Group Population & All & Radio\\
\hline
2 & 0.004 & 0.045\\
3 & 0.007 & 0.010\\
4 & 0.005 & 0.009\\
5 & 0.008 & 0.010\\
10 & 0.021 & 0.023\\
15 & 0.073 & 0.053\\
20 & 0.244 & 0.160\\
30 & 0.891 & 0.735\\
\hline
\hline
\label{monteradii}
\end{tabular}
\end{center}
\end{table}

From the Monte Carlo Shuffles we can see that the smaller radii associated with HT galaxy groups is statistically significant for groups with 10 or fewer galaxies. There does not appear to be a significant difference between the groups around radio galaxies and those around all galaxies. The slightly different values at the group population of two may be attributed to the smaller number of radio galaxies, and thus the likelihood of randomly selecting a HT galaxy group is higher.

\section{Discussion}
\subsection{Close Companions to the HT Galaxies}
Looking more closely at the high density grouping{s, we find that each HT galaxy has a close companion.
Using DSS images, we locate the closest galaxy as identified by the HIPASS group finder; these companion
galaxies are shown on radio-optical overlays in Figure~\ref{closecompanions}. Each HT galaxy has a close companion with projected separations ranging from 15 to 119 kpc.

\begin{figure*}
\begin{center}
\begin{tabular}{cc}
\includegraphics[scale=0.3]{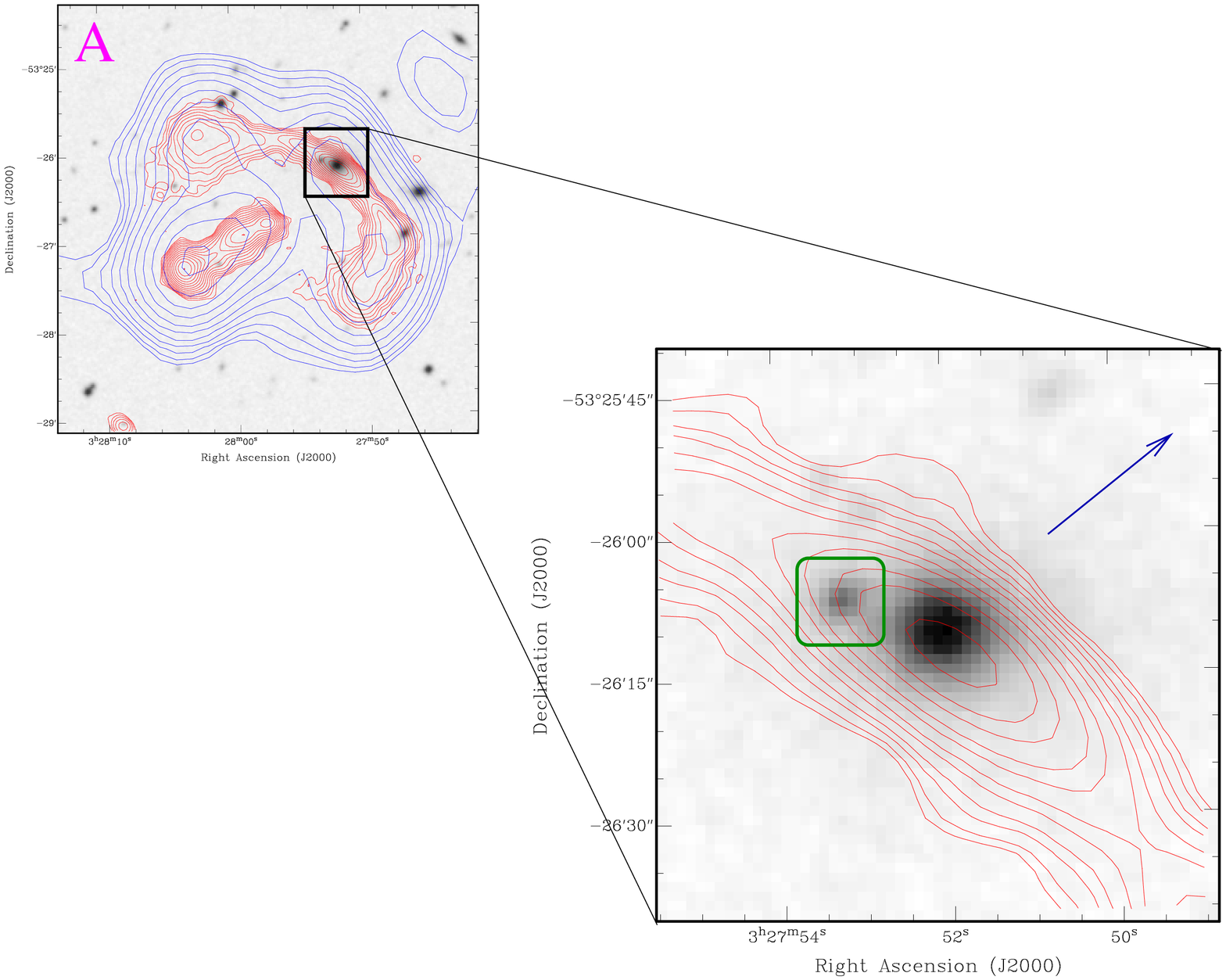} & \includegraphics[scale=0.3]{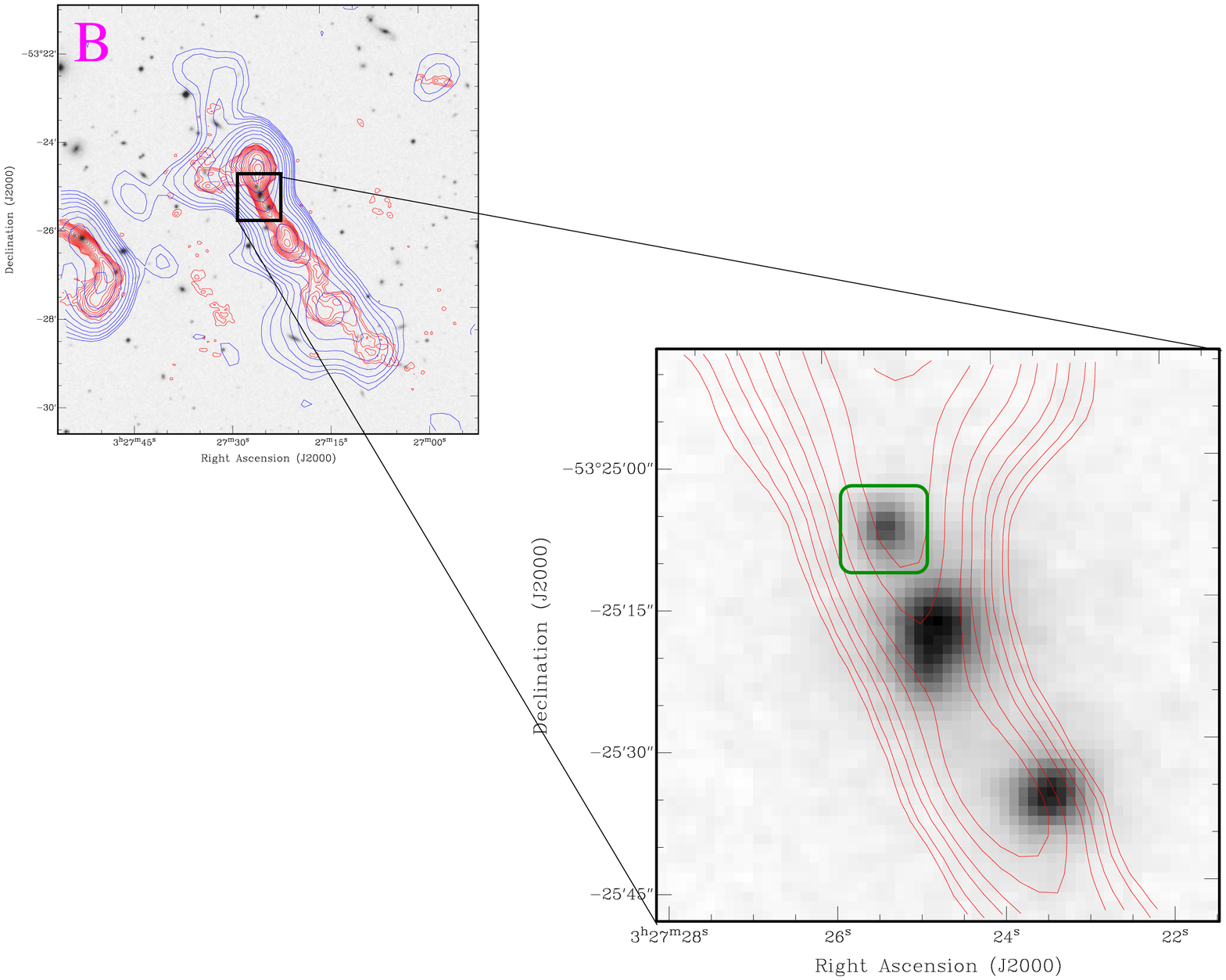}\\
\\
\includegraphics[scale=0.3]{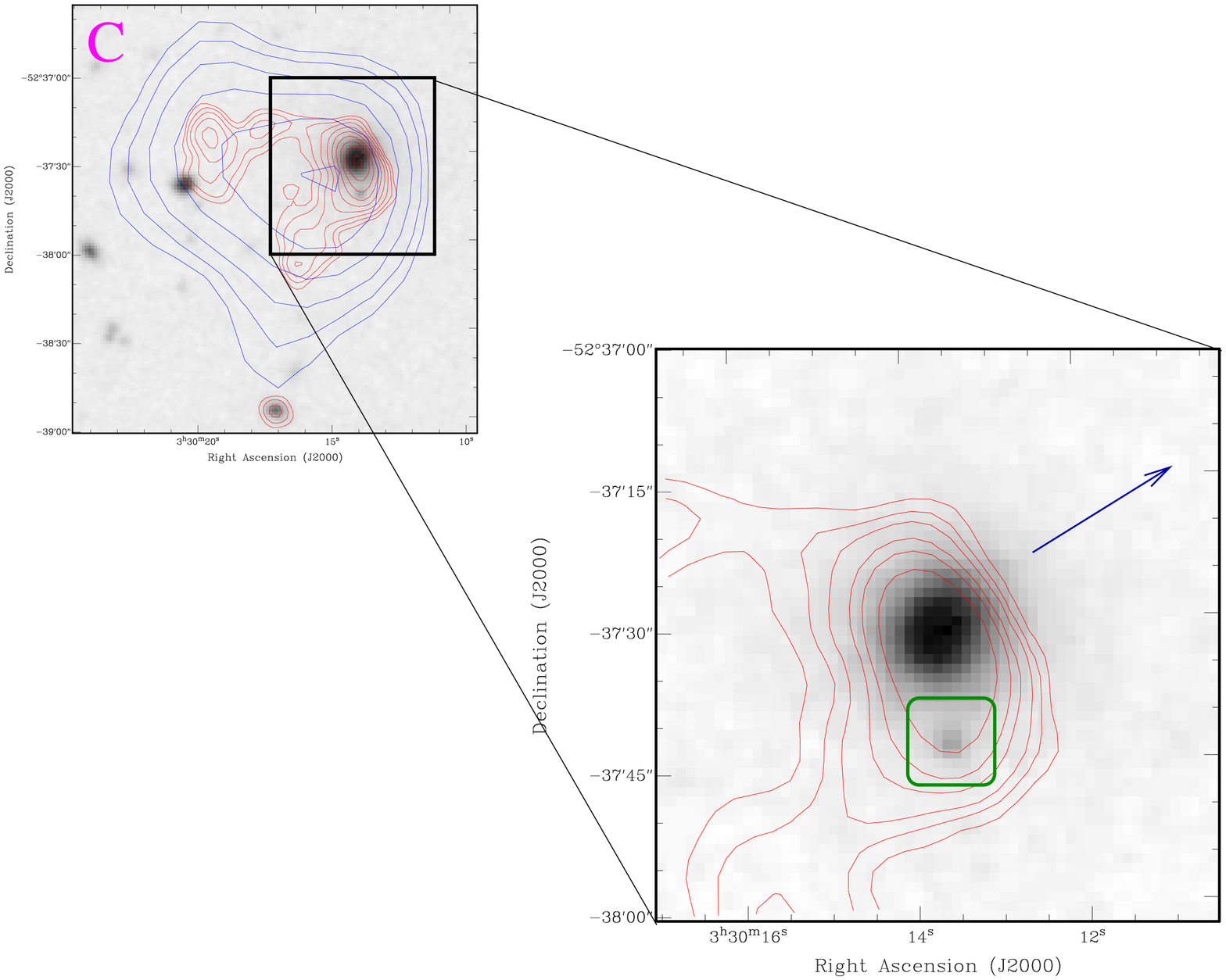} & \includegraphics[scale=0.3]{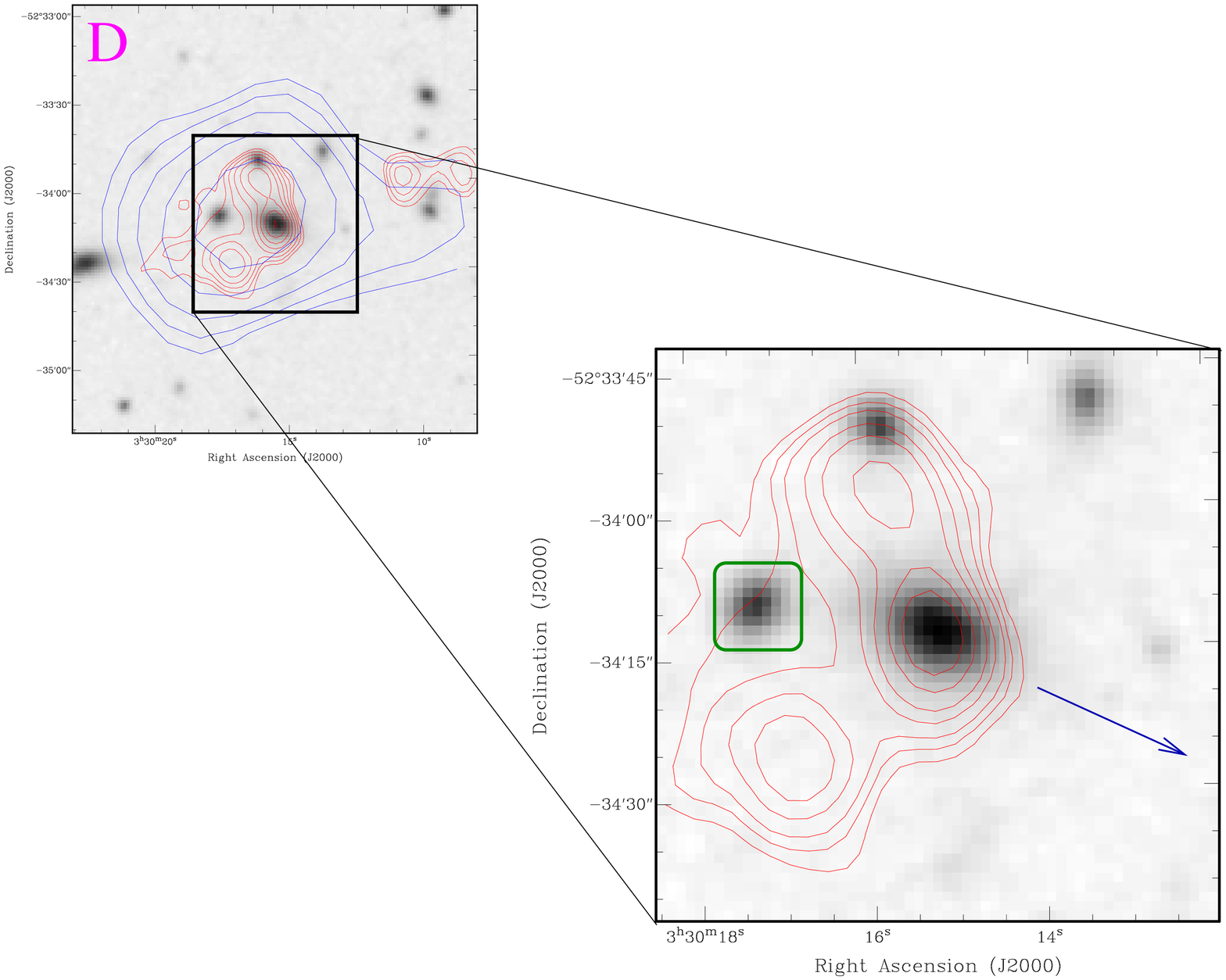}\\
\\
\includegraphics[scale=0.3]{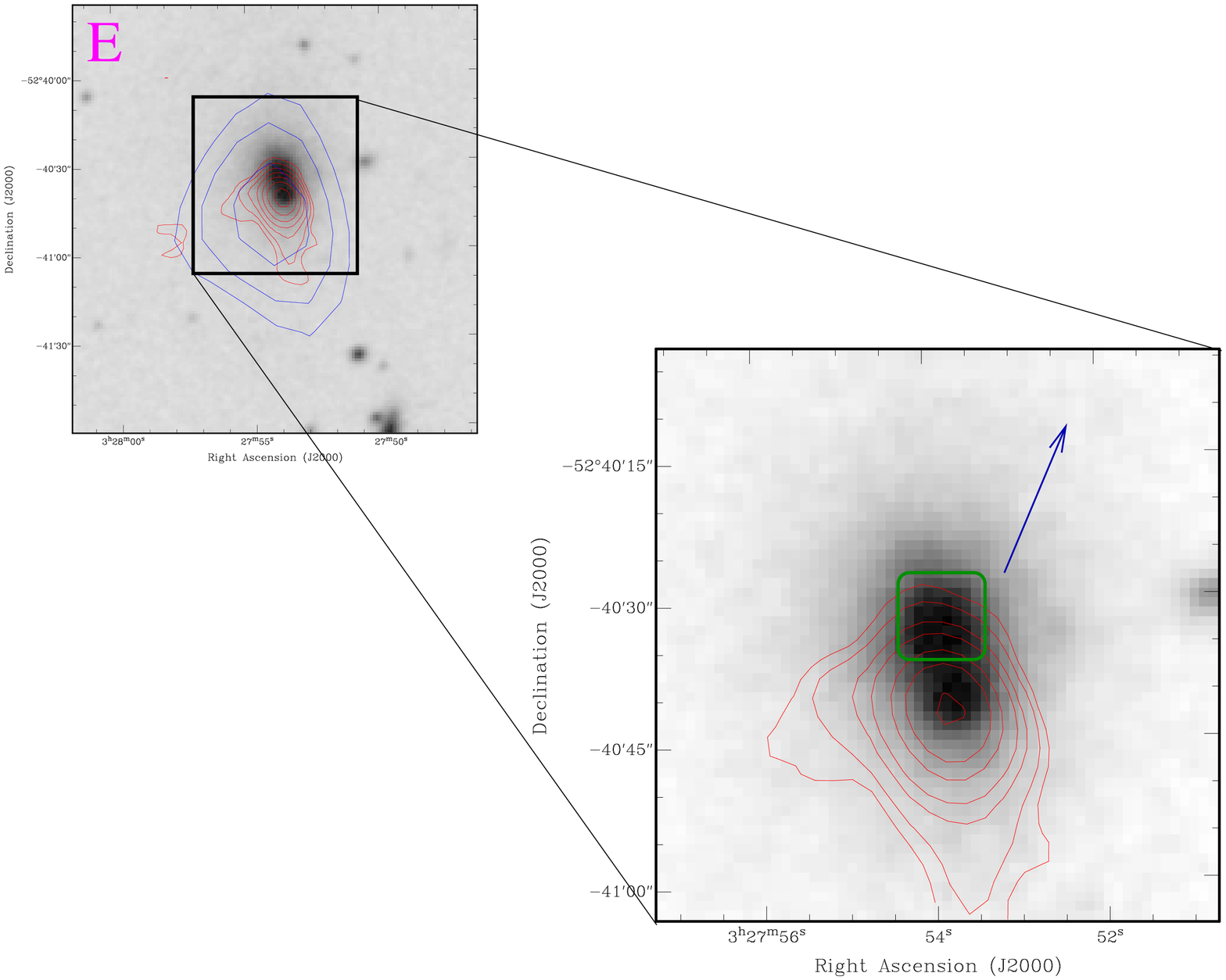}\\
\end{tabular}
\end{center}
\caption{The HT galaxies and their companion galaxies. The HT galaxy is shown in the ``zoomed out'' image while the optical galaxy and it's companion are shown in the ``zoomed in'' image. The 1.4 GHz ATCA data is red while the 843 MHz MOST data \citep{Mauch03} is blue. For both frequencies the lowest contour is three times the image RMS noise (0.1 mJy beam$^{-1}$ at 1.4 GHz) and contours increase at $\sqrt{2}$ intervals. As noted earlier, the radio galaxy nestled between the lobes of Galaxy A is a background source and not considered in this study. The green square identifies the closest companion to the HT galaxies as identified by eye. We note that this is the same as the companion identified using the HIPASS group finder in three cases. We do not have redshifts for the companion galaxies to Galaxies A and C - this is a possible explanation for their lower cut-off densities. The arrow indicates the hypothesized trajectory of the galaxy.}\label{closecompanions}
\end{figure*}

\subsection{Analysis of Statistics}
In our Results section we showed that the densities surrounding HT galaxies were significantly higher than those surrounding non-HT galaxies. In the following sections we will discuss the implications of this result. 

It has been well documented in the literature that there is no correlation between the density of a region and the number of radio galaxies observed \citep[e.g.][]{Venturi00}. Consequently the marginally significant P-values obtained for the non-radio v radio mean density test are startling as it suggests that radio galaxies exist preferentially in dense environments!

We are reluctant to believe that radio galaxies exist preferentially in dense environments. If anything, we expect the densities of the non-HT galaxies to be higher than the non-HT radio galaxies as high density non-radio galaxies could potentially skew our data. We attribute this bias to the difference in optical and radio coverage. The optical data have an average coverage of $\sim$90$\%$ down to a magnitude of 18. Conversely, our radio coverage is complete to flux densities of $\geq$ 0.5 mJy at 1.4 GHz. This is essentially 100\% complete for HT galaxies as the faintest HT galaxy has a flux density of 11 mJy at 1.4 GHz. The incomplete optical catalogue has the consequence of skewing the density of the optical catalogue downwards; that is, it appears less dense. 

From Figure \ref{skyplot} we can see that the optical coverage is 100\% down to m$_b$=18 over most of the field and the average coverage decreases uniformly over the field with fainter magnitude cutoffs. This means that the optical coverage surrounding the HT galaxies is not any different to the optical coverage surrounding the other radio galaxies. Thus, we think that the HT v non-HT test is free of optical biases.

The p-values obtained for both HT galaxy tests strongly suggest that the densities surrounding HT galaxies are significant and hence we are confident (to $\sim$99$\%$) that HT galaxies exist in significantly denser environments than non-HT galaxies.

The results derived from the 2D density enhancement analysis using projected radii appear to support the results we obtain from our 3D density enhancement study.

Our results also show that the tighter-knit environments surrounding HT galaxies remain significant for group populations of up to 10 galaxies. This suggests that high densities are required for HT galaxy morphology to occur.

\subsection{A Brief Foray into the Shapley Supercluster}
To ensure the density values obtained for the HT galaxies in the central region of the HRS are realistic, we test our process on the 
closest supercluster in the local universe: the Shapley supercluster. We do not have data for this region so we rely on the radio 
data available from \citet{Miller05} and \citet{Venturi00}. From their papers we identify five HT galaxies in the three central 
clusters of the Shapley supercluster. (The three clusters are a well-studied merger A3556-A3558-A3562, not dissimilar to A3125/A3128 
in the HRS.) The positional and redshift catalogue is taken straight from NED, so our visual catalogue compilation method is not 
utilized. The optical coverage of the Shapley supercluster is comparable to our data for the HRS \citep{Venturi00}. Furthermore, the 
sensitivity of the radio data is also comparable to our radio data \citep{Venturi00} so we will assume the two superclusters may 
be directly compared. The Shapley supercluster has roughly double the galaxy density of the HRS \citep{Proust06}. 

The density cut-off values for the HT galaxies in the Shapley supercluster are similar to the values obtained for the HT galaxies in the HRS (Table \ref{shapley}).

\begin{table}
\begin{center}
\caption{Cut-off densities for HT galaxies in Shapley.}\label{shapley}
\begin{tabular}{ll}
\\
\hline
\hline
& $V_l$ = 1000 km s$^{-1}$\\
HT galaxy & Cut-off density\\ & (galaxies/Mpc$^3$)\\
\hline
J132357-313845 & 8639\\
J133331-314058 & 1708\\
J132802-314521 & 183654\\
J133542-315354 & 367\\
J133048-314325 & 16729\\
\hline
\hline
\end{tabular}
\end{center}
\end{table}

The values for the Shapley supercluster are slightly higher than for the HRS but this may be attributed to the denser environment of the Shapley supercluster. However, the similarity of the results for a peculiar velocity of 1000 km s$^{-1}$ in both the HRS and the Shapley supercluster gives us confidence in the veracity of our results. Thus we may say with confidence that HT galaxies do form in regions of enhanced density.

\subsection{Further Evidence for High-Density Environments}
Inspection of archival ROSAT X-ray data shows that four of the five HT galaxies (A to D) reside in regions of enhanced emission near the centres of A3125 and A3128. There 
is also evidence that the fifth galaxy (galaxy E) is in a region of X-ray emission which is above the X-ray background but it is at a much 
lower level than for the other galaxies. Since X-ray emission also traces regions of enhanced density these results support the statistical analysis
presented above. Interestingly, \citet{Burns94} found a high correlation between X-ray emission and radio galaxies, and that X-ray emission around WATs was generally more extended. Furthermore, \citet{Miller99} found that X-ray emission around FRI galaxies is more extended than that around FRIIs. Our results are consistent with these findings. Details of the X-ray emission will be discussed in a separate paper.

The morphology of the HT galaxies is itself indicative of a dense environment,
as it is believed that the jets of FRI galaxies are less efficient in transporting energy due to the dense ICM surrounding them \citep{Fanaroff74}.

The morphology of HT galaxies has always been believed to be due to either:
\begin{itemize}
\item{the host galaxy having a high peculiar velocity (radio trail theory, \citep{Miley72}), or}
\item{the host galaxy being swept by strong winds (cluster weather, \citep{Burns98}).}
\end{itemize}

Using the optical data for our HT galaxies we find that the velocities are not exceptionally 
different to the cluster mean and hence it would appear that the host galaxies are victims of cluster 
weather. However, studies of cluster weather indicate that if cluster winds were 
the sole cause of the HT morphology, the velocity required for ram pressure is less than 
previously thought \citep{Pinkney95}. While we find that the velocities of the HT galaxies 
are not especially great, the densities in which they reside are among the densest regions of the 
cluster. Although we measure the number density of galaxies, this value follows
from the cluster potential, and thus should also be representative of the
density of the ICM due to gravitational attraction.

We propose that the morphology is due to ram pressure effects, but it is the exceptionally high gas densities that contribute to the morphology, as opposed to peculiar velocities or high velocity winds due to cluster weather (as P$_{ram}$$\propto$$\rho$$v^2$). 

\section{Conclusions}

We have compiled the largest positional and spectral catalogue of the major merger A3125/A3128. This well studied merger 
lies in the central region of the HRS, the second largest supercluster in the local universe. Using 1.4 GHz radio 
data we identified five HT galaxies in this region. 

To investigate the relationship between the HT galaxies and their underlying dynamical structure we identified 
substructure in the central region of the HRS. We found, using the HIPASS group finder, that 
the HT galaxies all exist in some of the densest regions of the cluster merger system A3125/A3128. We performed a series of statistical tests
to probe
three populations of galaxies within A3125/A3128 
(15,000 kms$^{-1}$ $\leq$ cz $\leq$ 35,000 kms$^{-1}$){\bf :} all galaxies (582), all radio galaxies with a
spectroscopic redshift within A3125/A3128 (46) and all HT galaxies (5). Results of the single-tailed permutation
tests showed, to 99$\%$ confidence, that HT galaxies reside in significantly denser environments than either non-HT
galaxies or radio galaxies which are known members of the clusters. These results are also replicated in a
comparative study of the five HT galaxies in the centre of the Shapley Supercluster.

Additionally, we found each of the HT galaxies in A3125/A3128 reside in regions of enhanced X-ray emission and have a very close companion (ranging from 15 - 119 kpc in projected separation). As
the peculiar velocities of the host galaxies are not significantly different to the mean cluster velocity this suggests that it is the density of the environment which plays a significant part in the bending of the jets. However, the presence of the close
companions is also likely to play a role in at least the motion of the host galaxies, though it is likely that 
the morphology of the jets on kpc-scales is more closely related to ram pressure and hence cluster weather. Thus, we propose that the combination of both a close companion and the cluster environment contribute to the bent morphology of HT galaxies.

Ultimately, we have found that HT galaxies are excellent indicators of dense regions and dynamical interactions within
galaxy clusters and thus are clear signposts of ``cluster weather''. As a result, HT galaxies may be used as 
barometers for the storms that occur in the cluster environment, although due to the complication of close 
companion galaxies, reading these barometers may not be as easy as first thought. We propose that the diffuse tails, on at least 
kpc-scales, can be used as giant wind socks and are effective beacons of cluster weather. Detection
of the full extent of HT galaxies requires low frequency observations of reasonable resolution. Instruments such as the GMRT and LOFAR will provide large samples of HT galaxies which may then be used as beacons of high-density regions in clusters.

\section*{Acknowledgments}

The detailed and extensive comments from our referee enabled us to clarify and vastly improve this paper. MM is grateful for the generous support of the Dave Warren Honours Scholarship in Optical Astronomy. This work was supported by the DFG cluster of excellence ``Origin and Structure of the Universe''. Many interesting and informative discussions with Ray Norris, Jim Lovell and Ron Ekers have also enriched this paper. The Australia Telescope Compact Array telescope is part of the Australia Telescope which is funded by the Commonwealth of Australia for operation as a National Facility managed by CSIRO. The Digitized Sky Survey was produced at the Space Telescope Science Institute under US Government grant NAG W-2166 and is based on photographic data obtained using The UK Schmidt Telescope. The UK Schmidt Telescope was operated by the Royal Observatory Edinburgh, with funding from the UK Science and Engineering Research Council, until 1988 June, and thereafter by the Anglo-Australian Observatory. Original plate material is copyright (c) of the Royal Observatory Edinburgh and the Anglo-Australian Observatory. The plates were processed into the present compressed digital form with their permission. This research has made use of the NASA/IPAC Extragalactic Database (NED) which is operated by the Jet Propulsion Laboratory, California Institute of Technology, under contract with the National Aeronautics and Space Administration. This research has also made use of NASA's Astrophysics Data System.

\label{lastpage}
\end{document}